\documentclass[aps,prl,twocolumn]{revtex4}
\usepackage{float}
 \usepackage{graphicx}
 \usepackage{subfigure}
\usepackage{amsfonts}
\usepackage{amssymb}
\usepackage{amsfonts}
\usepackage{dcolumn}
\usepackage{IEEEtrantools}
\usepackage{bm}
\usepackage{bbm}
\usepackage{amsmath}
\usepackage{amssymb}
\usepackage{amsbsy}
\usepackage{graphics}
\usepackage{makeidx}
\usepackage{graphicx}
\usepackage{appendix}
\usepackage[colorlinks,
linkcolor=blue,
anchorcolor=blue,
citecolor=blue]{hyperref}
\begin{document}
\title{Soliton sheets formed by interference of Bose-Einstein condensates in optical lattices }
\preprint{}
\author{Shu-Song Wang}
\affiliation{Institute of Theoretical Physics, State Key Laboratory of Quantum Optics and
	Quantum Optics Devices, Shanxi University, Taiyuan, Shanxi 030006, China}
\author{Su-Ying Zhang}
\email{zhangsy@sxu.edu.cn}
\affiliation{Institute of Theoretical Physics, State Key Laboratory of Quantum Optics and
	Quantum Optics Devices, Shanxi University, Taiyuan, Shanxi 030006, China}
\begin{abstract}
Soliton sheets which are formed by  interference of Bose Einstein condensates occupying  different single-particle states are observed in optical lattice potential. This structure consists of one-dimensional stationary solitons arranged periodically along the peaks of optical lattice (y direction) with the phase difference between the two sides of the soliton sheets is a linear function of y in each period, so we call it  soliton sheet. A y component velocity difference exists between the two sides of the soliton sheet. Similar velocity distributions can be produced by the alignment of an infinite number of isotropic vortices along the  peaks of the optical lattice. Their difference is that the soliton sheet structure is not limited by the number of phase singularities and can be generated even without phase singularities. 
\end{abstract}
\maketitle
\preprint{}
\affiliation{Institute of Theoretical Physics, State Key Laboratory of Quantum Optics and
	Quantum Optics Devices, Shanxi University, Taiyuan, Shanxi 030006, China}

\textit{Introduction.--} 
Topological excitations play an essential role in many areas of physics. There are various topological excitations  in Bose Einstein condensates(BECs),  like solitons\cite{sol V. V.Konotop,sol L.Salasnich,sol L.Salasnich W.B,sol V.Achilleos,sol V.E. Lobanov,sol Y. V. Kartashov,sol Y.-C. Zhang,sol Y.V.Kartashov,sol Y.Xu}, quantized vortices\cite{vor Siovitz ,vor A. Aftalion,vor B. Ramachandhran,vor T. Kawakami,vor X. F. Zhou,vor X.-Q. Xu}, vortex sheets\cite{Hanninen,  London, Landau, Onsager,sheet Han,sheet Kasamatsu,sheet parts}, domain walls\cite{dom B.A,dom L.E.Sadler}, textures\cite{tex C. McGarrigle,tex G. Ruben,tex M}. Topological excitations are essential to understand the phase, superfluidity and magnetic properties of condensates, such as spin textures are an intuitive response to the magnetic properties\cite{tex M} of multicomponent BECs, and solitons lead to phase differences\cite{sol V. V.Konotop} between the two ends of a one-dimensional condensate. Therefore the systematic study of topological excitations is essential.

The ground state structure of BECs in a rotating system is influenced by the shape of the external potential.  BECs in a harmonic  potential  respond to rotation by producing an Abrikosov triangular vortex lattice\cite{A J. R. Abo-Shaeer}. Giant vortices\cite{giant A.C.White} can be produced in  rotating BECs confined in  a shifted harmonic potential, and hidden vortices\cite{hidden T. Mithun,hidden} are present in BECs in the double-well potential.

In this letter, we study the ground state properties of a rotating  BEC in   optical lattice. We find a soliton sheet structure that is formed by the interference of condensates occupying different single-particle states having different y component of velocities. We conduct a comprehensive analysis of the density, phase, and velocity properties exhibited by these soliton sheets. There is local period density modulation of the condensate along the soliton sheet. At each density period, the phase difference between the two sides of the soliton sheet is a linear function of y. Along the centerline (peak of the optical lattice) of the soliton sheet, the velocity can be described by the tangent of half of the phase difference  for the x component and a constant  for the y component. The velocity y(x) component on both sides of the soliton sheet can be expressed by the single-particle velocity y(x) component together with a small fluctuation, which can be described by the cosine (sine) of the phase difference and decreases as the distance  from the condensate to the soliton sheets  increases, with the fluctuation tending to zero when the distance is sufficiently far.

\textit{Single-particle ground state.---}We consider a  two-dimensional (2D)  BEC in a rotating frame in a one-dimensional (1D) optical lattice. The single-particle Hamiltonian $\hat H_{0}$ can be given by  
\begin{equation}
\hat{H}_{0}= -\frac{{\hbar ^{2}}}{{%
2M}}{\nabla ^{2}}+V_{OL}+V_{H}-\Omega {\hat{L}_{z}}   \label{H0},
\end{equation}%
where $
\hat{L}_z=-i{\hbar }\left( x\partial _y-y\partial _x \right) $ is the $z$ component of the angular momentum operator.
 The external potential
in this work consists of two parts, the 2D harmonic trapping potential $V_{H}=%
\dfrac{1}{2}M \omega _{\bot }^{2}\left[\left( x^{2}+y^{2}\right) \right] $
and the optical lattice potential $V_{OL}=V_0\cos^2\left( \kappa x \right) $. Here $\kappa $ is the wave vector of the laser used to generate the optical lattice potential.
Stationary solutions of single-particle systems with  the eigenenergy $ E $ are sought in the usual form
\begin{equation}
\Psi \left( x,y,t \right) =\psi \left( x,y \right) \exp \left( -i Et/\hbar \right) \label{fenli},
\end{equation}%
with stationary function $\psi $ satisfying the equation
\begin{equation}
\left( -\frac{{\hbar ^{2}}}{{%
		2M}}{\nabla ^{2}}+V_{OL}+V_{H}-\Omega {\hat{L}_{z}} \right) \psi =E\psi \label{dingtai}.
\end{equation}%
 We introduce a generalized momentum operator $\hat{P}=\left( \hat{P}_x,\hat{P}_y \right) $, where
 \begin{IEEEeqnarray}{rCl}
 	\hat{P}_x=-i{\hbar }\partial _x+M\varOmega y,
 	\IEEEyesnumber \IEEEyessubnumber \label{px} \\  
 	\hat{P}_y=-i{\hbar }\partial _y-M\varOmega x,
 	\IEEEyessubnumber    \label{py}
 \end{IEEEeqnarray}
and $\hat H_0 $ can be rewritten as
\begin{equation}
 \hat H_0=\frac{1}{2M}\hat{P}_{x}^{2}+\frac{1}{2M}\hat{P}_{y}^{2}+V_{OL}+\frac{1}{2}M\left( \omega_{\perp} ^2-\varOmega ^2 \right) \left( x^2+y^2 \right) \label{H0var}.
\end{equation}%
We find the single-partcle state in two cases:\newline
(i) $\Omega=\omega _{\perp }$ .
The Eq(\ref{H0var})can be simplified as
\begin{equation}
	\hat H_0=\frac{1}{2m}\hat{P}_{x}^{2}+\frac{1}{2}\hat{P}_{y}^{2}+V_{OL} \label{H0var1}. 
\end{equation}%
Obviously,  $\hat{P}_{y}$  commutes with the single-particle Hamiltonian $\hat H_0$, i.e., $\left[ \hat{P_y},\hat{H}_{0}\right] =0$.
Thus $\hat{P_y}$ and $\hat{H}_{0}$ have common eigenstates. The eigenequation
of the $\hat{P_y}$ operator can be expressed as%
\begin{equation}
\hat{P}_y\psi ^m=k_{y}^{m}\psi ^m,  \label{pyzhi}
\end{equation}%
where  $ \psi^{m} $ ($ m=0, \pm1 \text{,} \pm 2\text{,...} $) labels the eigenstate of $\hat{P_y}$, and $ k_{y}^{m} $ denotes the eigenvalue of the operator $\hat{P_y}$ in the
eigenstate $\psi^m $. Solving Eq(\ref{pyzhi}), we obtain the eigenstate  as%
\begin{equation}
\psi ^m\left( x,y \right)=u(x)\exp\left(i\theta^m(x,y)\right) , \label{bianfeng}
\end{equation}
where $\theta^m(x,y)=k_{y}^{m} y/\hbar-M\Omega xy/\hbar$.
Substituting Eq.(\ref{bianfeng}) into Eq.(\ref{H0var}), we get
\begin{eqnarray}
(-\frac{{\hbar ^{2}}}{{%
		2M}}\partial _{x}^{2}+V_{eff}) u^{n,m}=E^{n,m} u^{n,m}, \label{1ddingtai}
\end{eqnarray}%
where $V_{eff}=2M\Omega^2\left(  x-x^m_t \right) ^2+V_0\cos^2 \left( \kappa x \right)$ represents the effective potential that determines the  single-particle density distribution,  $x^m_t=mT $ labels the horizontal coordinate of the troughs of optical lattice with $T=\pi/\kappa$ denoting the optical lattice period, and $u^{n,m}$ labels the eigenstate of $\hat{H_0}$. Obviously, the phase of $u^{n,m}$ is  space-independent. Thus, ${\left| u^{n,m} \right|}^2$ and $\theta^m$  describe the single-particle  density distribution and  phase distribution, respectively. All single-particle energy eigenstates can be labeled with two quantum numbers $\left| n,m \right> $. Within a given $m$ sector, the lowest-energy eigenstate will be assigned $n = 1$. Since we only consider $n = 1$ in this letter, this quantum number is omitted in the next representation.
To find the single-particle ground state, we construct the  energy functional corresponding to Eq.(\ref{1ddingtai}) as
\begin{eqnarray}
	E^m&	=&\int {d}\mathbf{r}u^{m\dag}(-\frac{{\hbar ^{2}}}{{%
			2m}}\partial _{x}^{2}+V_{eff})u^m.
	 \label{energy}
\end{eqnarray}%
  Using this ansatz to minimize the energy, it is easy to find 
\begin{eqnarray}
	k_{y}^{m}=2M\varOmega x^m_t.  \label{pyquzhi}
\end{eqnarray}%
 Different  $m$ correspond to distinct single-particle degenerate states corresponding to the same energy. The Eq.(\ref{1ddingtai}), a one-dimensional Schrödinger equation of x, shows that along the y-direction, consistent with the classical case, the centrifugal and external harmonic trap cancel each other, and the single-particle density distribution is y-independent. However, along the x-direction,  unlike the classical case, the centrifugal and external harmonic trap cancel each other but an additional secondary potential $ 2M\Omega^2\left(x -x^m_t \right) ^2 $ is introduced. The term $ 2M\Omega^2\left(x -x^m_t \right) ^2 $  allows the single-particle density  $\left| m \right> $ is mainly bound in the optical lattice of the m-th period.
\newline
(ii) $\Omega\ne \omega _{\perp }$. The phase distribution obtained in case (i) can be generalized to  $\Omega\ne\omega_{\perp}$. We  introduce a phase correction factor  $\gamma_m$ based on the results of case (i) and assume that $\psi^{m} \left( x,y \right) =u^{m}\left( x,y \right) \exp \left( i\left( k_y^m y/\hbar-\gamma_mM\varOmega xy/\hbar \right) \right)$.  By substituting this expression into Eq.(\ref{dingtai}), and ignoring the spatial derivative term of $\gamma_m$, we obtain
\begin{eqnarray}
&&-i\hbar\left( k_y^m -\left( 1+\gamma_m \right)M\varOmega x \right)\partial_yu^m-i\hbar\left( 1-\gamma_m \right)M\varOmega y\partial_xu^m \notag \label{Hxiu}
\\&& -\frac{\hbar^2}{2M}\partial_{x}^{2}u^m-\frac{\hbar^2}{2M}\partial_{y}^{2}u^m+V_{x}u^m +V_{y}u^m =E^mu^m, 
\end{eqnarray}%
where $V_{x}=\dfrac{1}{2}M\left( \omega _{\bot}^{2}-\varOmega ^2 \right) x^2 +2M\Omega^2(x-x_t^m)^2+V_0\cos^2(\kappa x)
 $ and $V_y=\dfrac{1}{2}M\left( \omega _{\bot}^{2}-\varOmega ^2 \right) y^2$ represent the effective external potential in the x and y directions, respectively. The degeneracy of single-particle states is disrupted by the
\begin{figure}[t]
	\centering
	\includegraphics[trim=0 180 10 80,clip,
	width=0.95\linewidth]{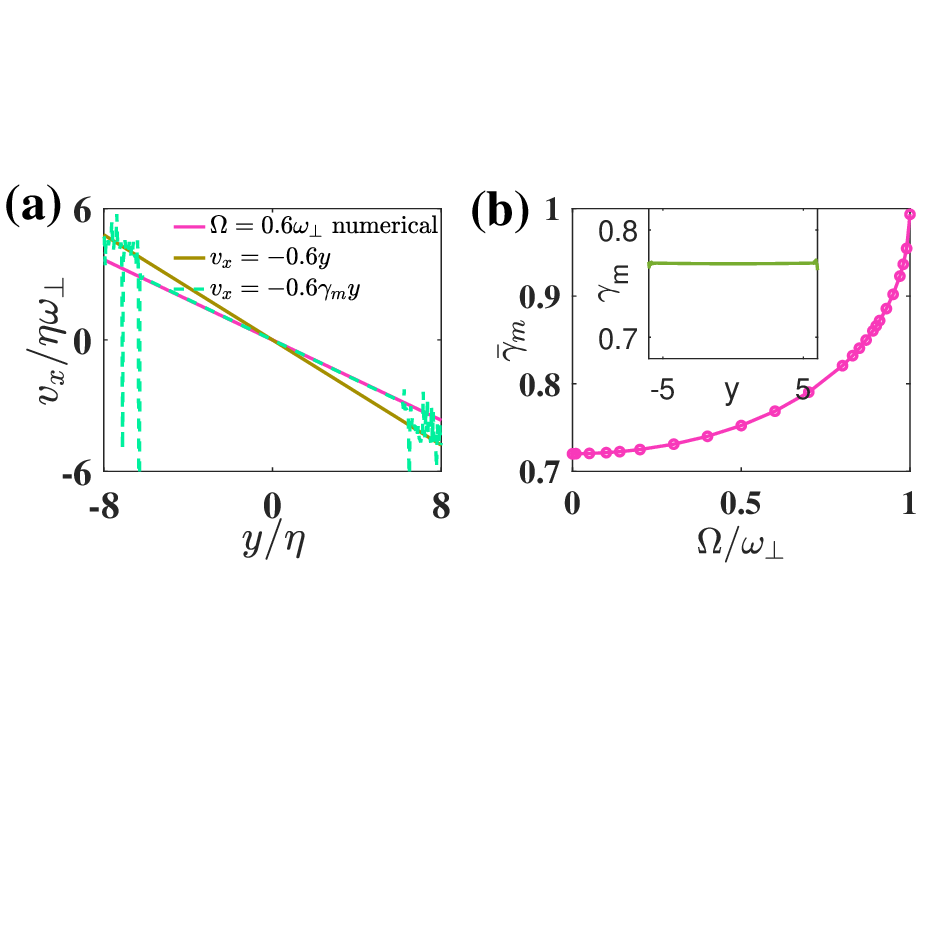}%
	\caption{(a) Section views of  $v_x$ are presented along the $y$ axis. The numerical results deviate significantly from  $v_x=-\Omega y$, but show good agreement with $v_x=-\Omega \gamma_m y$, which represents the y-component of the velocity after incorporating the correction factor, where the rotating angular frequency $\Omega=0.6\omega_{\bot}$ and the period of the optical lattice potential $T=\pi\eta$ with $\eta=(\hbar/m\omega_{\bot})^{1/2}$. (b) $\bar{\gamma}_m$ exhibits a positive relationship with $\Omega$, and specifically, $\gamma_m=1$ when $\Omega=\omega_{\perp}$. The inset demonstrates that $\gamma_m$ can be approximated as a constant for a given $\Omega$. 
	} \label{danlizisudu}
\end{figure}
presence of $\dfrac{1}{2}M\left( \omega _{\bot}^{2}-\varOmega ^2 \right) x^2 $, and leads to large $m$ corresponding to large single-particle energies. 
Since $\hat{H}_0$ is a Hermitian operator, we obtain 
\begin{equation}
\left( 1-\gamma_m \right)M\varOmega y\partial_xu^m+\left( k_y^m -\left( 1+\gamma_m \right)M\varOmega x \right) \partial_yu^m=0 \label{xiuzheng}.                  \end{equation}
Specifically, when $x = x^m_t$, the single-particle density takes a relative maximum along the $x$ direction, with $\partial _xu^m=0$. According to Eq.(\ref{gamma}) we get $k_y^m=\left( 1+\gamma_m \right) \varOmega x^m_t$. When $y=0$, the single particle
 density takes a relative maximum along the $y$ direction, with $\partial _yu^m=0$ and Eq.(\ref{xiuzheng}) naturally holds whatever $\gamma_m$ takes. When $y\ne 0$ and $x\ne x^m_t$, substituting $k_y^m=\left( 1+\gamma_m \right) \varOmega x^m_t$ into Eq.(\ref{gamma}), we get 
\begin{eqnarray}
 \gamma_m =\dfrac{ \partial _xu^m/\left(x-x^m_t\right)-\partial _yu^m/y }{\partial _xu^m/\left(x-x^m_t\right)+\partial _yu^m/y} \label{gamma}.
\end{eqnarray}
According to Eq(\ref{gamma}), we can calculate the  x component of the single-particle velocity containing the correction factor $\gamma_m$ is 
\begin{eqnarray}
	v_x =-\gamma_m \Omega y \label{vxxiuzheng}.
\end{eqnarray}
As shown in Fig.\ref{danlizisudu}(a), the corrected x-component of the velocity is in perfect agreement with the numerical results.
According to Eq.(\ref{Hxiu}), the single particle is confined in an anisotropic effective potential with the binding in the x-direction is significantly stronger than in the y-direction, resulting in $\partial_xu^m/\left( x-x^m_t \right)\gg \partial_yu^m/y $. For a given $\Omega$,  $\gamma_m$ can be approximated as a constant that is  less than 1. Hence, it is reasonable to neglect the spatial derivative term of $\gamma_m$ in Eq.(\ref{Hxiu}).
For varying $\Omega$, the large $\Omega$ corresponds to a weak  bound along the y-direction, since $V_{y}=\dfrac{1}{2}M\left( \omega _{\bot}^{2}-\varOmega ^2 \right) y^2  
$. Therefore $\gamma_m$ increases as $\Omega$ increases, as shown in Fig.\ref{danlizisudu}(b).
Specifically, when $\Omega=\omega _{\perp }$, the density is homogeneously distributed along the $y$ direction with $\partial_yu^m=0$ and from Eq.(\ref{gamma}) we get $\gamma_m=1$ , which is consistent with the result we obtained in case (i).

\textit{Condensate ground state.---}In this subsection, we explore the ground state of the condensate containing repulsive interactions. The interaction Hamiltonian takes the form
\begin{eqnarray}
	H_{int}=\frac{1}{2}\int{dr^2} \beta \rho ^2,
	\label{Hint}
\end{eqnarray}%
where $\rho =\left| \psi \right|^2$ is the condensate density, which obeys the normalization condition $\int{dr^2}\rho \left( \boldsymbol{r} \right) =1$, $ \beta $ characterizes the inter-particle interaction strength which can be tuned by optical Feshbach resonances. The
total Hamiltonian $H$ of the condensate is given by $H=H_0+H_{int}$. We use analytical calculations and numerical simulations to study the condensate ground state, and in the following discussion we fix the optical lattice potential  depth $V_0=20 \eta$ with $\eta = \sqrt{{\hbar }/M\omega _{\bot}} $, optical lattice period $T=\pi \eta$ and the interparticle interaction strength $\beta=1000 \hbar^2/M$. The solid line in the figure represents the numerical simulation results, and the dashed line represents the analytical results.

Single-particle states are essential for understanding condensate ground state properties. Similar to the discussion of single-particle states we analyze the condensate ground state in two cases:\newline
(i) $\varOmega =\omega _{\bot}$. 
 We assume that the condensate wave function is a linear superposition of the single-particle states
\begin{eqnarray}
	\psi =\sum_m{\sqrt{N_m}\left| m \right> }=\sum_m{\sqrt{N_m}u^m\exp \left( i\theta ^m \right) }
	\label{zuhe},
\end{eqnarray}%
where $N_m $, $u^m$ obeys the normalization condition $\sum_m{N_m}=1$  and $\int{dr^2}|u^m|^2=1$ . The $u^m$ is required to satisfy
\begin{equation}
(-\frac{\hbar^2}{2m}\partial _{x}^{2}+V_{eff}+\beta N_m\left| u^m \right|^2) u^{m}=E u^{m}. \label{gdingtai}
\end{equation}%
where $V_{eff}=2M\Omega( x- x^m_t ) ^2+V_0\cos^2 ( kx+\varphi )$ represents the effective potential. Condensates in $\left| m \right> $ predominantly occupy the optical lattice of the m-th period, and the interaction energy between condensates in different single-particle states is extremely small.
Using this ansatz to minimize the interaction energy,  it favors $N_{m1}=N_{m2}$ $(m1\ne m2) $.

\begin{figure}[htb]
	\centering
	\includegraphics[trim=5 170 10 40,clip,
	width=0.95\linewidth]{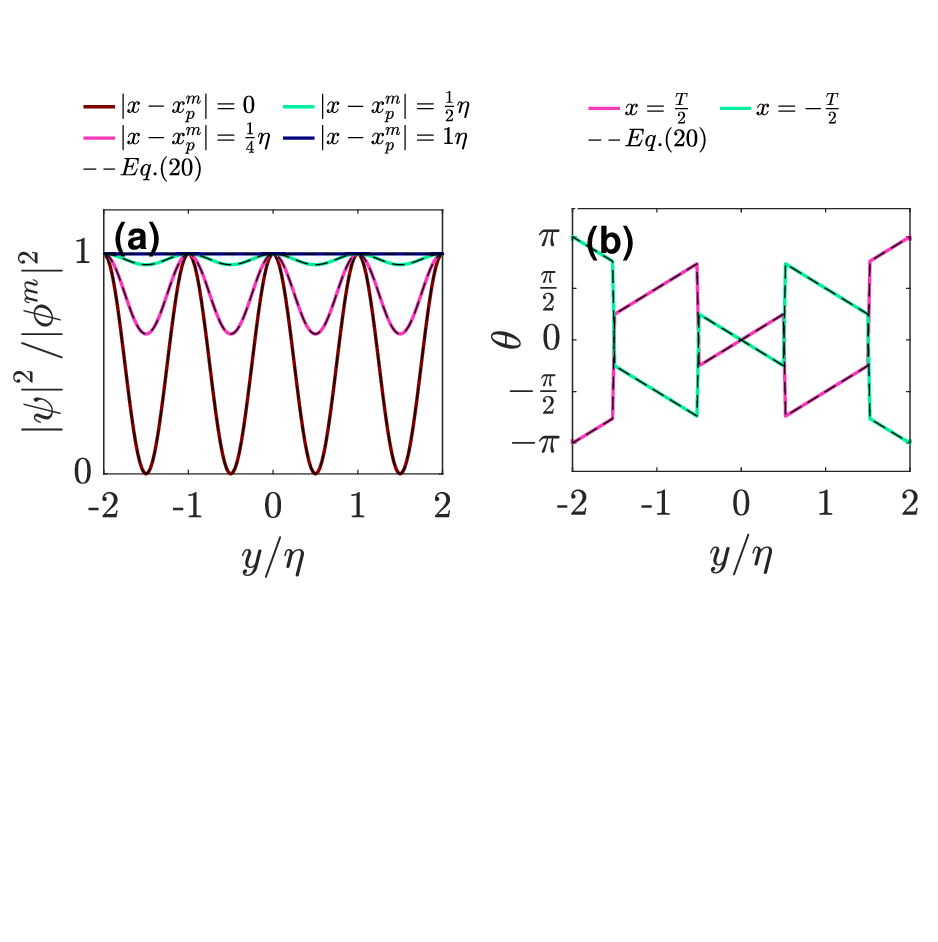} %
	\caption{(a) Density cross section for condensate interference for $\Omega=\omega_{\bot}$. The interference effect is most pronounced when $x = x^m_p$. As the distance from the condenser to the optical lattice peak increases, the interference effect decreases significantly. When $|x-x^m_p|/\eta>$1, the interference effect is almost negligible. (b) Section views of phase $\theta$  along $x=T/2$ and $x=-T/2$. The  $\theta$ changes in a linear manner with respect to $y$ and has a $\pi$ phase jump at $y=(k+\frac{1}{2})\eta$ ($k$ is an integer), corresponding to the point with $|\varphi|^2= 0$ in figure (a). } \label{center}
\end{figure}

For simplicity, we first consider the superposition of two adjacent single-particle state wave functions and give the following expression
\begin{eqnarray}
	\psi^{m} &=&\sqrt{N_m}\left| m \right> +\sqrt{N_{m+1}} \left| m+1 \right>  \notag\\
	 &=&\phi^{m}
	(\alpha\cos \frac{\varDelta  \theta}{2}  +(1-\alpha)  \exp ( i\frac{\varDelta \theta}{2} ) )
	\label{interf}.
\end{eqnarray}%
where $\phi^{m}=\sqrt{N_m} (u^m+u^{m+1})	\exp ( i( \theta_m +\frac{\varDelta\theta}{2})) $, $\alpha=\dfrac{2u^m}{u^m+u^{m+1}}$ and $\varDelta \theta =\theta ^{m+1}-\theta ^{m}=2M\varOmega Ty/\hbar$. We obtain that the condensate density can be expressed as
\begin{eqnarray}
	|\psi^{m}|^2 =|\phi^m|^2(1-a(2-a)\sin^2\frac{\varDelta  \theta}{2} )
	\label{interfden}.
\end{eqnarray}%
 Around the peak of the optical lattice, the  densities of the condensate in $\left| {m} \right>$ and $\left| {m+1} \right>$ are comparable, resulting in a noticeable interference effect. The interference breaks the continuous translational symmetry  and leads to localized  periodic density modulation with period $\hbar\pi/MT\Omega$ along the optical lattice peaks. The interference effect is most pronounced on the peaks of  optical lattice  precisely with $a=1$. 	The period density modulation effect becomes invisible rapidly when there is a significant disparity between the condensate density in $\left|m \right>$ and $\left| m+1 \right>$, observed on both sides of the optical lattice peak. When $|x-x^m_p|/\eta>$1 with $\alpha\rightarrow 0$ or $\alpha\rightarrow 2$, the interference effect the interference effect can be neglected, there is $\left| \varphi \right|^2=1$, as shown in Fig.\ref{center}(a), where $x_p^m=x_t^m+T/2$ labels the horizontal coordinate of the  peaks optical lattice.
		
Along the peaks of the optical lattice,  $\psi^m$ can be simplified as 
\begin{equation}
	\psi^m(x_p^m,y) =2\sqrt{N_m}u^m\cos ( \frac{\varDelta\theta}{2} ) \exp (i(\theta_m+\frac{\varDelta\theta}{2}) )  
	\label{linjie}.
\end{equation}
We get
	\begin{eqnarray}
		\varphi( x^m_p, y_{z} +0^+ )=-\varphi ( x^m_p,y_{z} +0^- ),
		\label{qidian}
	\end{eqnarray}%
where $y_{z}=( \dfrac{1}{2}+k ) \hbar\pi /MT\Omega$ $(k=0,\pm 1,...)$ . Consequently,  there is a $\pi$ phase difference across these points, as show in Fig.\ref{center}(b). 
	
In order to study intuitively the ground state structural properties of the condensate, $\psi^m$ can be reexpressed as
\begin{eqnarray}
	\psi^{m} &=& \sqrt{N_m}\left( u^{m}+u^{m+1} \right) \exp \left( i\theta ^{m+1} \right)\varphi^{m^-}
	\notag \\ &=&\sqrt{N_m}\left( u^{m}+u^{m+1} \right) \exp \left( i\theta ^m \right)\varphi^{m^+}, \label{phim}
\end{eqnarray}%
where
\begin{IEEEeqnarray}{rCl}
		\varphi^{m^-}&=& \frac{1+\exp \left( -i \varDelta\theta   \right)u^{m}/u^{m+1}}{1+u^{m}/u^{m+1}}	,
	\IEEEyesnumber \IEEEyessubnumber \label{vphijian} \\  
 \varphi^{m^+}&=& \dfrac{1+\exp \left( i\varDelta \theta   \right)u^{m+1}/u^{m}}{1+u^{m+1}/u^m}.
 \IEEEyessubnumber    \label{vphijia}
\end{IEEEeqnarray}
\begin{figure}[t]
	\centering
	\includegraphics[trim=10 95 0 160,clip, width=0.95\linewidth]{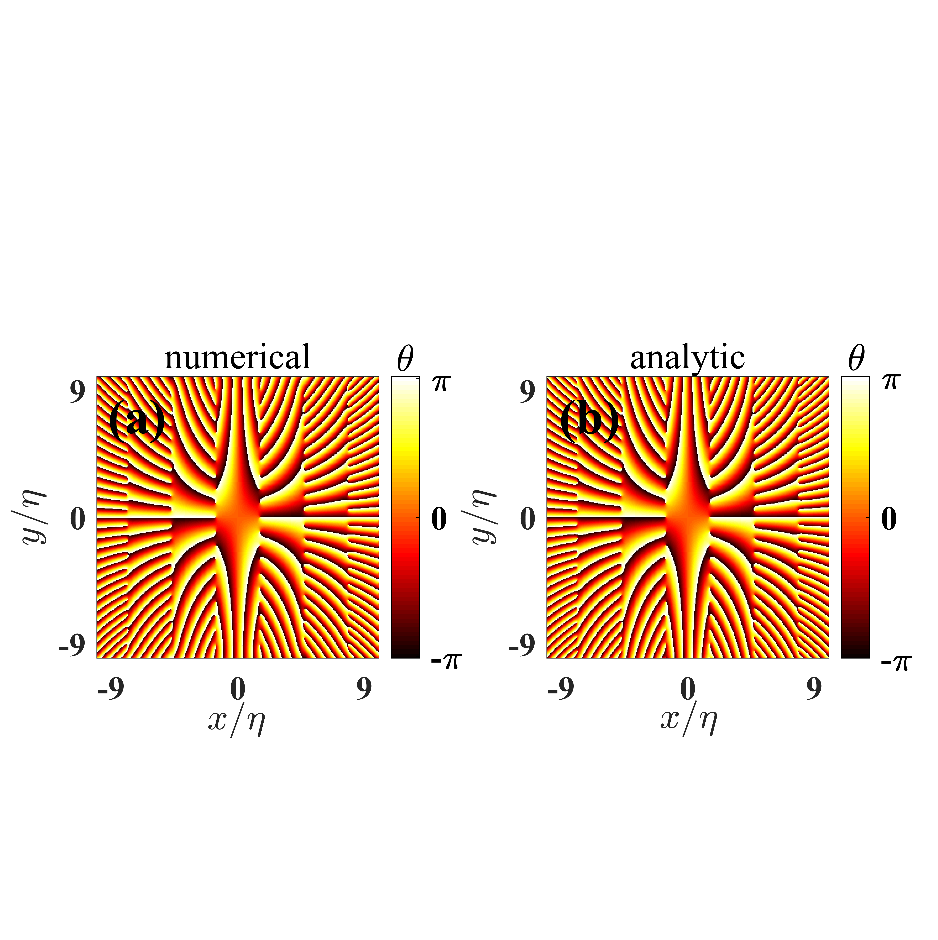} %
	\caption{ The condensate phase distribution for $\Omega=\omega_{\bot}$, with the numerical results on the left column and the analytical results on the right column are in perfect agreement.} \label{zongxiangwei}
\end{figure}
For a given y, the density and phase characteristics described by Eq. (\ref{vphijian}) and Eq. (\ref{vphijia}) align with the one-dimensional soliton. This indicates that the one-dimensional solitons are arranged along the peaks of optical lattice and form  a new topological excitation, which we call the soliton sheet.
 Without loss of generality, Eq.(\ref{vphijian}) and Eq.(\ref{vphijia}) can be approximately rewritten as
 \begin{IEEEeqnarray}{rCl}
 	\varphi^{m^-}&=& \frac{1+\exp({\left( x^m_{p}-x \right) /\xi}) \exp \left( -i \varDelta\theta   \right)}{1+\exp({\left( x^m_{p}-x \right) /\xi}) }	,
 	\IEEEyesnumber \IEEEyessubnumber \label{vphijian2} \\  
 	\varphi^{m^+}&=& \dfrac{1+\exp({\left( x-x^m_t \right) /\xi}) \exp \left( i\varDelta \theta   \right)}{1+\exp({\left( x-x^m_p \right) /\xi}) },
 	\IEEEyessubnumber    \label{vphijia2}
 \end{IEEEeqnarray}
where $\xi$ denotes the width of the soliton sheets, which is determined by potential depth  $V_0$ of the lattices and the strength of the inter-particle interaction $\beta $.

Consider that condensates in distinct single-particle states do not spatially overlap except for adjacent single-particle states, the condensate wave function can be expressed as
\begin{eqnarray}
	\psi =\sum_m{\sqrt{N_m}\left| {m} \right>}=\psi '\prod_{m< 0}{\varphi ^{m^-}}\prod_{m\ge 0}{\varphi ^{m^+}} ,  \label{sum}
\end{eqnarray}%
where 
\begin{eqnarray}
	\psi '=\sum_m  \sqrt{N_m}u^{m} \exp( -i\Omega xy ), \label{backgroud}
\end{eqnarray}%
which we call the background wave function.
Fig.\ref{zongxiangwei} illustrates the phase distribution of the condensate, demonstrating a full alignment between the numerical and analytical results. 

\begin{figure}[htb]
	\centering
	\includegraphics[trim=0 70 0 25,clip, width=0.95\linewidth]{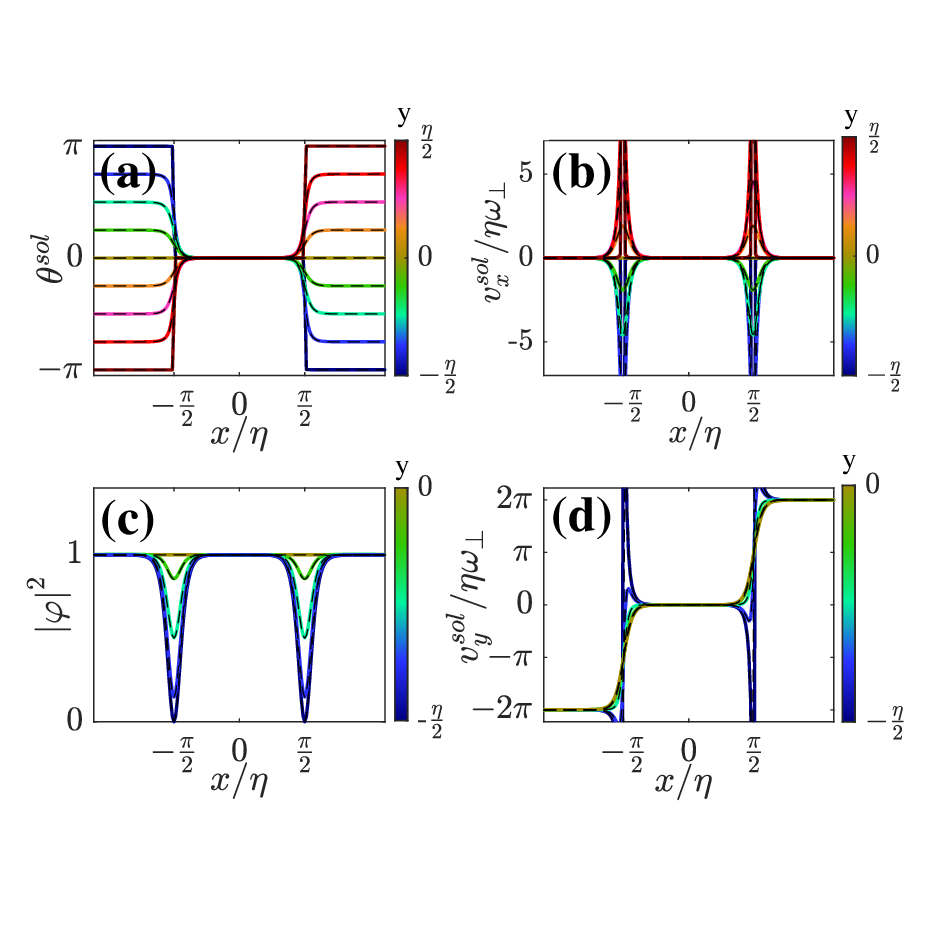} %
	\caption{(a) Cross-section of the phase distribution for one period, where $\Omega=\omega_{\bot}$. There is a $\varDelta\theta$ phase difference between the two sides of the soliton sheet. The $y$ varies $\hbar\pi/M\Omega T$ corresponding to one period of $\varDelta\theta$, which   ranges from $-\pi$ to $\pi$. (b) There exists a local $x$ component $v_x$ velocity along the soliton sheet. There is a correspondence between $v_x$ and $\varDelta\theta$, with larger $\varDelta\theta$ corresponding to larger $v_x$.  When $\varDelta\theta=\pi$ or $\varDelta\theta=-\pi$, $v_x$ tends to infinity. (c)Since $y=\pm y_0$ correspond to the same density distribution, we  give the cross section of the density distribution for half a period. As the value of $|v_x^{sol}|$ increases, the local densities decrease, and when $|v_x^{sol}|$ tends to infinity, the density is 0. (d) The cross section of the distribution of the $y$ component $v_y^{sol}$ of the velocity. There is a constant y-component velocity difference between the two sides of the soliton sheet. }. \label{soliton fig}
\end{figure}
To enhance our comprehension of the characteristics exhibited by the soliton sheet, we shall exclude the background wave function and instead consider
\begin{eqnarray}
   \varphi=\psi/\psi '=\prod_{m < 0}{\varphi ^{m^-}}\prod_{m\ge 0}{\varphi ^{m^+}}
	\label{soliton sheet}.
\end{eqnarray}%
Obviously, this wave function has the following properties that $\lim_{x\rightarrow x^m_t}\theta^{sol} =m\varDelta \theta $ and $\lim_{x\rightarrow x_{m+1}}\theta^{sol} =\left( m+1 \right) \varDelta \theta $, where $\theta^{sol} $ is the phase of $\varphi$.
So we find that that there is a $\varDelta\theta$ phase difference between the two sides of the soliton sheet, resulting in a localized rapid phase change on the soliton sheet. Localized rapid phase changes result in localized x component $v^{sol}_x$ of velocity, with greater phase jumps correspond to greater $v^{sol}_x$ and $v^{sol}_x$ tends to infinity when the phase difference is $\pm \pi$. Since the  velocity x-component is zero in $\left| m \right> $ and $\left| m+1 \right>$ without considering the background wave function. Eq.(\ref{soliton sheet}) suggests that the superposition of condensates in different single-particle states with zero  x-component of velocity excites a locally non-zero  x-component of velocity. From the energy point of view, the local non-zero velocity inevitably leads to the decrease of the local density. When $v^{sol}_x$ tends to infinity, the local density tends to zero, as show in Fig.\ref{soliton fig}(a),(b),(c). Meanwhile, since $\varDelta\theta $ is a function of y, we can calculate the y-component velocity difference between the two sides of the soliton sheet is
\begin{eqnarray}
	\varDelta v_y=\hbar\partial _y\varDelta \theta /M=2\Omega T
	\label{deltav},
\end{eqnarray}%
as shown in Fig.\ref{soliton fig}(d).
\begin{figure}[htb]
	\centering
	\includegraphics[trim=0 80 0 120,clip,
	width=0.95\linewidth]{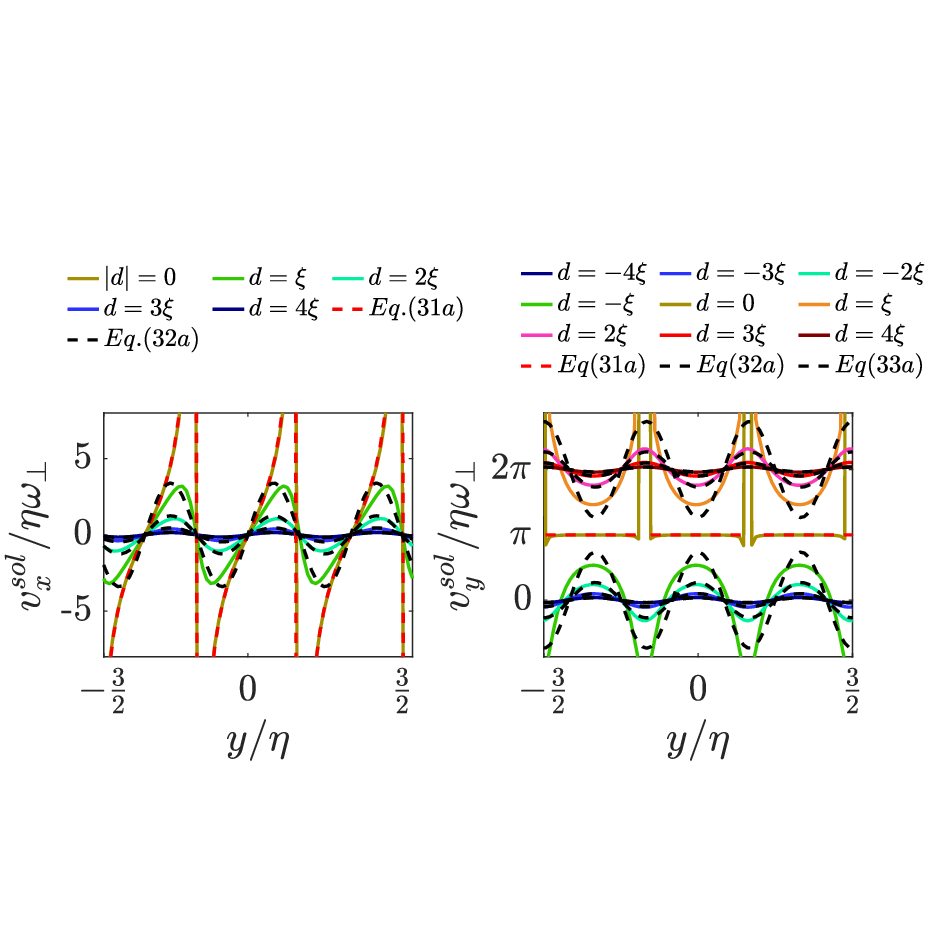} %
	\caption{ Section views of  $v_x$, $v_y$ are presented along the $y$ axis. The variable $d=x-x^1_p$ denotes the distance from the condensate to the peak  of optical lattice . For $|d|>2\xi$, the numerical simulation results are in good agreement with Eq.(31) and Eq.(32). For $|d|>4\xi$, the $x$ component of the velocity is approximately zero and the $y$ component  is approximated to be a constant. } \label{vfig}
\end{figure}

To deeply investigate the superfluid properties of the condensate, we compute the x-component and y-component of the velocity except for the phase singularity as 
 \begin{subequations}
 	\begin{flalign}
 		 v_x^{sol}=&\sum_m{\frac{\hbar}{2\xi M}\frac{\chi \sin \left( \varDelta\theta \right)}{1 +2\chi \cos \left( \varDelta\theta \right)+\chi^2}
 		 } ,\label{vxsol}\\
 		v_y^{sol}=&\sum_{m\ge 0}{\varDelta v_y \frac{\cos \left( \varDelta\theta \right) \chi +\chi^2}{1 +2\chi \cos \left( \varDelta\theta \right)+\chi^2}} \notag \\ 
 		+&\sum_{m<0}{-\varDelta v_y \frac{1+\cos \left( \varDelta\theta \right) \chi}{1 +2\chi \cos \left( \varDelta\theta \right)+\chi^2}}
 		,\label{vysol}
 	\end{flalign}
 \end{subequations}
where $\chi=\exp(\frac{x-x^m_p}{\xi})$. Specifically, along the centerline of the soliton sheet with $x=x^{m}_p$ ,
 we get
\begin{IEEEeqnarray}{rCl}
	v_x^{sol}( x^m_p,y )&=&\frac{\hbar}{2\xi M}\tan \frac{\varDelta\theta }{2},
	\IEEEyesnumber \IEEEyessubnumber 
	\label{vx} \\  
	v_y^{sol}( x^m_p,y )&=&(m+\frac{1}{2})\varDelta v_y
	.
	\IEEEyessubnumber  
	\label{vy}
\end{IEEEeqnarray}
When $|\varDelta\theta| = \pi$, the $x$ component of velocity tends to infinity, corresponding to a phase singularity. As shown in Fig. \ref{vfig} the numerical results are in perfect agreement with Eq.(31).

When $x\rightarrow x^{m}_t$, ignoring higher order terms of $ \chi $, we get
\begin{subequations}
	\begin{flalign}
v_x^{sol}\left( x\rightarrow x^m_t \right) =&\frac{\hbar}{\xi M}\chi
 \sin \varDelta\theta ,
	\IEEEyesnumber \IEEEyessubnumber 
	\label{vxsolm+1} \\  
v_y^{sol}\left( x\rightarrow x^m_t \right) =&m\varDelta v_{y}+\chi \varDelta v_{y}  \cos {\varDelta\theta} 	.
	\IEEEyessubnumber  
	\label{vysolm+1}
\end{flalign}
\end{subequations}
Similarly, when $x\rightarrow x^{m+1}_t$, there are
\begin{subequations}
	\begin{flalign}
		v_x^{sol}\left( x\rightarrow x_{m+1} \right) &=\frac{\hbar}{\xi M}\frac{1}{\chi}
		  \sin \varDelta\theta ,\label{vxsolm}\\
		v_y^{sol}( x\rightarrow x_{m+1} ) &=(m+1)\varDelta v_{y}-\frac{1}{\chi} \varDelta v_{y}  \cos {\varDelta\theta} . \label{vysolm+1}	
	\end{flalign}
\end{subequations}
 This shows that the x (y) component of velocity  on both sides of the soliton sheet can be expressed as a constant ( for the x component the constant is 0) together with a small fluctuation. Fig.\ref{vfig} shows that when $d>2\xi$, the approximation of Eq.(32) and Eq.(33) are valid, and when $d>4\xi$ the, fluctuations tend to 0 resulting in the velocity x(y) component can be approximated as a constant,  where $d=x-x_p^m$.
 
Consider the velocity distribution generated by a vortex array with the same vorticity located at $(0,nl)$, which can be expressed as \cite{zh}
\begin{IEEEeqnarray}{rCl}
v _x^{vor} &=&-\frac{\hbar}{l M}\sum_{n=-\infty}^{+\infty}{\frac{y/l+n}{\left( y/l+n \right) ^2+\left( x/l \right) ^2}},
	\IEEEyesnumber \IEEEyessubnumber 
	\label{vxvor} \\  
	v _y^{vor} &=&\frac{\hbar}{l M}\sum_{n=-\infty}^{+\infty}{\frac{x/l}{\left( y/l+n \right) ^2+\left( x/l \right) ^2}},
	\IEEEyessubnumber  
	\label{vyvor}
\end{IEEEeqnarray}
where $l$ represents the distance between adjacent vortices. When $x>l$  there are $v_x\rightarrow 0$ and $v_y\rightarrow \hbar\pi/lM$ and  when $x<-l$  there are $v_x\rightarrow 0$ and $v_y\rightarrow -\hbar\pi/lM$. The velocity distribution resulting from an array of infinite vortices located at $(0,nl)$ shares similar characteristics with the velocity distribution observed on both sides of the soliton sheets. The vortex array is characterized by a scale of $l$ , while the soliton sheet is characterized by a scale of $\xi$. When the distance between the condensate and the peaks of the optical lattice exceeds the characteristic scale, the x component of  velocity tends to zero, while the y component of  velocity tends to a constant. Additionally, there is a velocity difference in the y component between the two sides of the  peak of optical lattice. In contrast to vortex arrays, which necessitate an infinite number of phase singularities to achieve this velocity distribution, soliton sheets are not constrained by the number of phase singularities. The soliton sheet  provides a new explanation for the domain wall structure in Ref. \cite{zh}. Additionally,  the velocity on both sides of the domain wall is consistent with the single-particle velocity, aligning with the viewpoint presented in Ref. \cite{zh}.
 
 According to the effect of the soliton sheet on the velocity distribution and combining with Eq. (\ref{sum}), the velocity distribution of the condensate is easily obtained as
\begin{IEEEeqnarray}{rCl}
v_x=v_{x}^{sol}-\Omega y,
	\IEEEyesnumber \IEEEyessubnumber 
	\label{vxact} \\  
v_y=v_{y}^{sol}-\Omega x,
	\IEEEyessubnumber  
	\label{vyact}
\end{IEEEeqnarray}
where $\Omega x$, $\Omega y$ are the contributions of the background wave function to the velocity distribution.
\begin{figure}[t]
	\centering
	\includegraphics[trim=0 175 10 80,clip,
	width=0.95\linewidth]{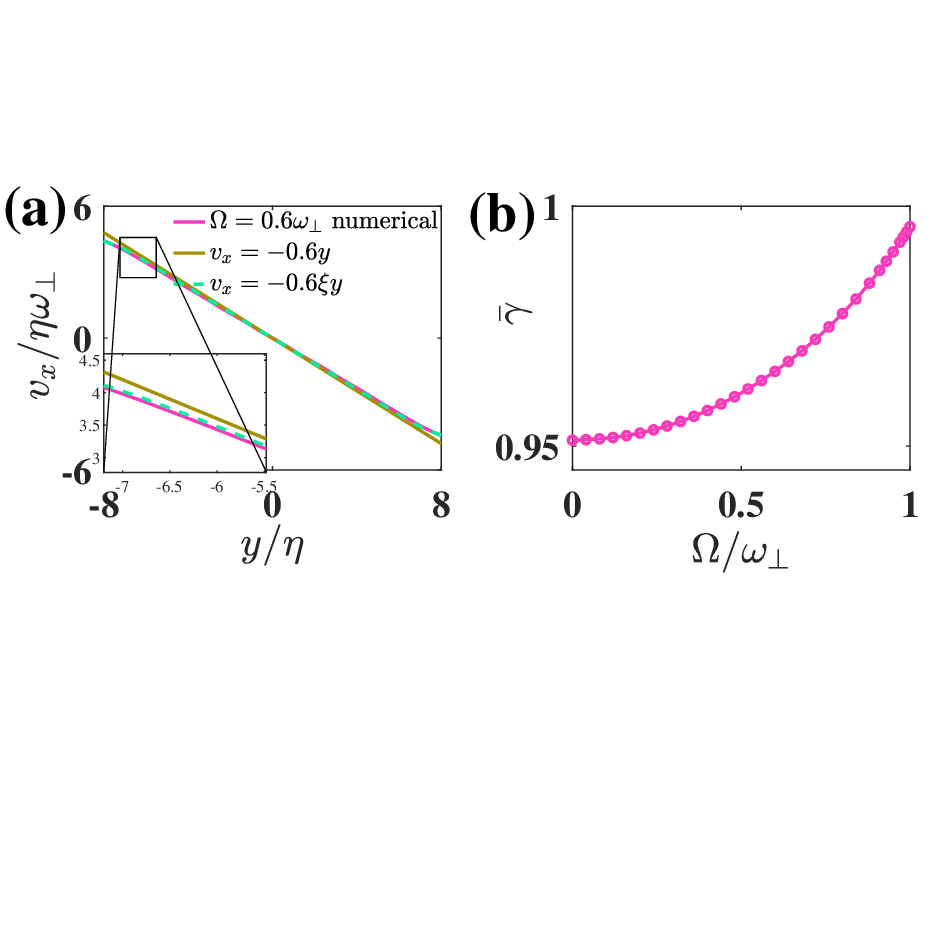} %
	\caption{ Section views of  $v_x$ are presented along the $y$ axis. The local zoom plots show that the numerical results deviate slightly from $v_x=-\Omega y$ but are in almost perfect agreement with $v_x=-\Omega \gamma y$, which represents the y-component of the velocity after the addition of the correction factor, where $\Omega=0.6\omega_{\bot}$. (b) $\bar{\gamma}$ exhibits a positive relationship with $\Omega$, with $\bar{\gamma}$ always greater than 0.95 even when $\Omega$ is small. } \label{conxiuzheng}
\end{figure}
\newline
(ii) $\varOmega \ne \omega _{\bot}$. Similarly to case (i), the condensate ground state is still  described by Eq.(\ref{zuhe}) , with the difference that single-particle state degeneracy is lifted due to the presence of $\dfrac{1}{2}M\left( \omega _{\bot}^{2}-\varOmega ^2 \right) \left( x^2+y^2 \right) $, and the smaller $|m|$ is, the lower the single-particle state energy is. The condensate tends to occupy the lowest energy state until the inter-particle interaction energy exceeds the energy level  difference between   the adjacent single particle states, resulting in  $N_{m1}>N_{m2}$ when $m_1<m_2$.

When the interaction is large enough, the spatial scale of the condensate distribution along the y-direction expands considerably compared to that in the single-particle state, resulting in a small $\partial_yu$. According to Eq.(\ref{gamma}), $\bar\gamma$ will be significantly closer to 1 compared to the single-particle state, as shown in Fig.\ref{conxiuzheng}, $\gamma$ is always greater than 0.95, while Fig. \ref{danlizisudu} shows that in the single-particle state, $\gamma_m$ is close to 0.7, when $\Omega$ is small enough. 

The superposition of single-particle states also interferes and produces soliton sheets, but the center of the soliton sheets are no longer precisely centered at the optical lattice peaks , instead being in near of it. 
\begin{figure}[tph]
	\centering
	\includegraphics[trim=10 90 0 155,clip, width=0.95\linewidth] {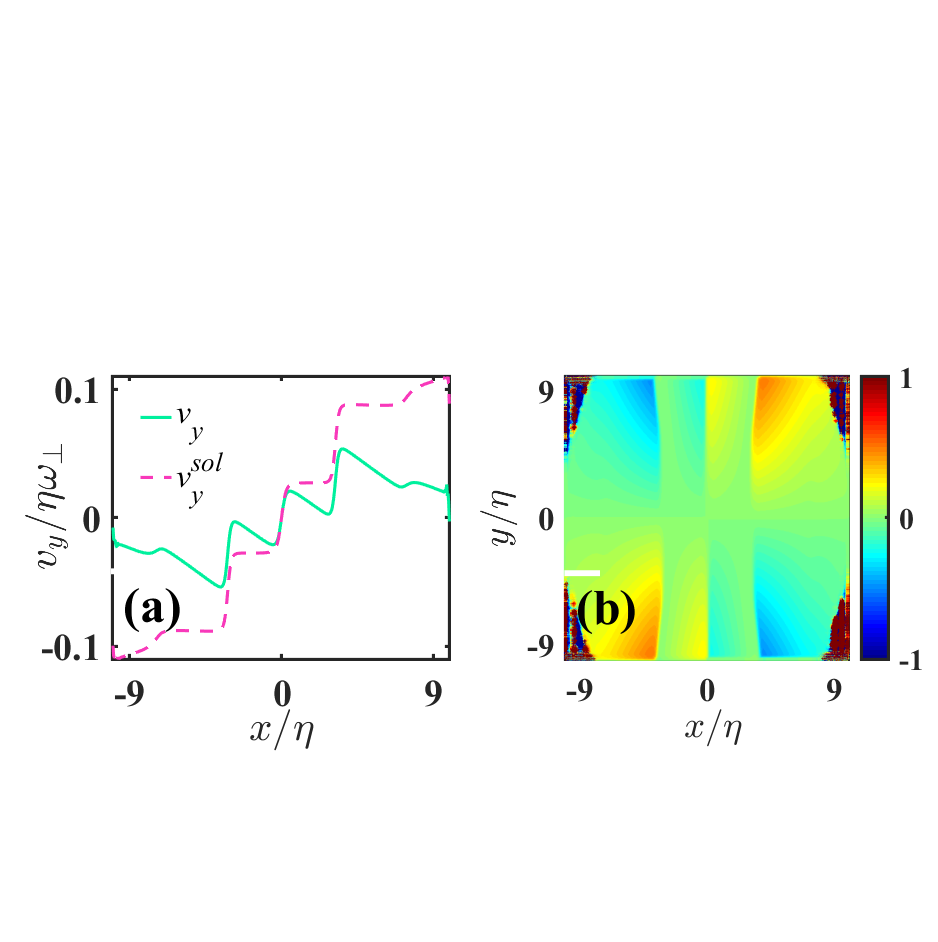} %
	\caption{(a)Section views of  $v_y$ are presented along the $x$ axis. The box marks the location of the y-component velocity jump, indicating that a soliton sheet was created in the condensate. (b) The condensate phase distribution indicates that no vortices were produced in the condensate.} \label{0.01}
\end{figure}

The period of spatial density modulation due to vortex slices should be corrected to be $\pi/\gamma M T\Omega$. At the center of the soliton sheet, a phase singularity exists every $\pi/\gamma MT\Omega$. When $\Omega$ is sufficiently small, this period can be larger than the scale of the condensate distribution along the y-direction. This results in the presence of soliton sheets but not vortices in the condensate. The results of the numerical simulations for $\Omega=0.01\omega_{\bot}$ are shown in Fig.\ref{0.01}. The tangential velocity jumps suggest that soliton sheets are produced in the condensate, while the phase distribution indicates the absence of vortex   in the condensate. 
According to Onsager-Feynman quantization condition
 \begin{eqnarray}
\oint_{C}{\mathbf{v}}\cdot d\mathbf{l}=\frac{2\pi \hbar }{M}N,
	\label{OF}
\end{eqnarray}%
Considering that the y-component of the single-particle velocity is y-independent, if we choose the two sides of the sheet as the integration path and assume that the region enclosed by the path does not contain phase singularities, we get that the y-component velocity difference $\varDelta v_y $ between the two sides of the sheet is
\begin{eqnarray}
	\varDelta v_y=2\Omega T.
	\label{deltav}
\end{eqnarray}
This suggests that a y-dependent phase difference leads to a velocity difference on two sides of the soliton sheets, and that phase singularities are not essential for the generation of this structure.

\textit{Conclusion.---}  We have investigated the soliton sheet structure, which is formed by the interference of condensates in different single-particle states, in a rotating BEC in a one-dimensional optical lattice potential. The single-particle state is solved in two cases:(i)  $\Omega=\omega _{\perp }$, the condensate density is uniformly distributed along the $y$ direction.  The single-particle phase $\theta^m(x,y)=M\Omega y(2x^m_t-x) /\hbar$ can be obtained analytically by defining the generalized momentum operator. (ii) $\Omega\neq\omega _{\perp }$, the condensate density is non-uniformly distributed along the $y$ direction. We generalize the phase distribution for case (i) to $\Omega\neq\omega _{\perp }$ by introducing a correction factor $\gamma_m$ . The $\gamma_m$ can be determined by substituting the corrected phase into the Hamiltonian function and verifying that the eigenvalues of the Hamiltonian are real. The corrected velocity distribution is in perfect agreement with the numerical results.

The condensate ground state can be represented as a superposition of single-particle states which interferes near the  optical lattice peaks. We find that the condensate density can be expressed by a sine function. This is due  to the fact that the interference disrupts the continuous translational symmetry in the y-direction leading to a  periodic density modulation with period $\pi/\gamma MT\Omega$  and forms soliton sheets.  The phase difference between the two sides of the soliton sheet is  $\varDelta\theta$.  The  $\varDelta\theta$ ranges from $-\pi$ to $\pi$ and changes in a linear manner with respect to y  in each density period. The phase differences between the two sides of the soliton sheet lead to rapid phase changes and localized  velocities around the sheet. The expression of the velocity distribution  is provided, revealing that  when $\varDelta\theta=\pi$ or $\varDelta \theta=-\pi$, the local velocity tends to infinity, indicating the presence of a phase singularity. The distance between two phase singulars on a soliton sheet is $\pi/\gamma MT\Omega$, which increases as $\Omega$ decreases. When  $\Omega$ is sufficiently small and $\pi/\gamma MT\Omega$ exceeds the size of the condensate distribution  along the $y$ direction, there are soliton sheets but there  exist no vortices in the condensate.

This work is supported by National Natural Science Foundation of China (11772177) and Fundamental Research Program of Shanxi Province (202203021211310).

\end{document}